%% file: mainArxive.tex
\documentclass[letterpaper, 10 pt, conference]{ieeeconf}

\IEEEoverridecommandlockouts                              

\usepackage{cite}
\usepackage{amsmath,amssymb,amsfonts}
\usepackage{algorithmic}
\usepackage{graphicx}
\usepackage{algorithm,algorithmic}
\usepackage[hidelinks]{hyperref}

\usepackage{comment}

\allowdisplaybreaks 

\usepackage{pgfplots}
  \usepgfplotslibrary{groupplots}
\usepgfplotslibrary{fillbetween}
 \pgfplotsset{compat=newest}

\usepackage{subcaption}

\usepackage{tikz}
\usetikzlibrary{positioning, arrows, calc, shadings}

\tikzstyle{blockdiag}	= [node distance=5mm, >=stealth', semithick]
\tikzstyle{block}			= [draw, rectangle, minimum width=1cm, minimum 
height=.8cm]
\tikzstyle{sum} = [draw,circle,inner sep=0pt, minimum size=6pt]
\tikzstyle{connector} = [draw,circle,inner sep=0pt, minimum size=2pt, 
fill=black]

\usepackage{array}
\newcommand{\norm}[1]{\left\|#1\right\|}
\newcommand{\abs}[1]{\left|#1\right|}
\newcommand{\field}[1]{\mathbb{#1}}
\newcommand{\R}{\field{R}}

\newcommand{\Rn}{\field{R}_{\ge 0}}

\newcommand{\Se}{\mathcal{G}}

\newcommand{\Mcmp}{\mathcal{M}_\text{inc}}
\newcommand{\Minc}{\mathcal{M}_\text{inc}}
\newcommand{\Sym}{\field{S}}
\newcommand{\N}{\field{N}}
\newcommand{\M}{\mathcal{M}}
\newcommand{\Md}{\M_{\text{d}}}
\newcommand{\Mfb}{\M_{\text{fb}}}
\newcommand{\Mfc}{\M_{\text{c}}}
\newcommand{\D}{\mathcal{D}}
\newcommand{\diag}[1]{\text{diag}(#1)}
\newcommand{\Inc}{\mathcal{I}}
\newcommand{\QC}[1]{\mathcal{QC}(#1)}

\newcommand{\bmtx}{\begin{bmatrix}}
\newcommand{\emtx}{\end{bmatrix}}
\newcommand{\bsmtx}{\left[ \begin{smallmatrix}} 
\newcommand{\esmtx}{\end{smallmatrix} \right]} 
\newcommand{\bmatarray}[1]{\left[\begin{array}{#1}}
\newcommand{\ematarray}{\end{array}\right]} 

\definecolor{blue1}{RGB}{222,235,247}
\definecolor{blue2}{RGB}{158,202,225}
\definecolor{blue3}{RGB}{49,130,189}

\newtheorem{mytheo}{Theorem}

\newtheorem{defn}{Definition}

\newtheorem{lem}{Lemma}
\newtheorem{cor}{Corollary}

\newcommand{\edtn}[1]{\null} 

\newcommand{\fb}[1]{{\color{orange}{(FB: #1)}}}

\begin{document}


\title{An Exact, Finite Dimensional Representation \\ for Full-Block, Circle Criterion Multipliers}

\author{Felix Biert\"umpfel,
Bin Hu,
Geir Dullerud, 
Peter Seiler
\thanks{F. Biert\"umpfel  and P. Seiler are with the Department of 
Electrical Engineering \& Computer Science, at the University of 
Michigan ({\tt\small felixb@umich.edu;} and
		{\tt\small pseiler@umich.edu}).}
        \thanks{F. Biert\"umpfel is also with the Chair of Flight Mechanics and Control at TU Dresden ({\tt \small felix.biertuempfel@tu-dresden.de})}
\thanks{B. Hu is with the Department of Electrical and Computer 
Engineering at the University of Illinois at Urbana-Champaign 
({\tt \small binhu7@illinois.edu}).}
\thanks{G. Dullerud is with the Department of Electrical and Computer Engineering at the University of Minnesota, Twin Cities
({\tt \small dullerud@umn.edu}).}
}

\maketitle

\begin{abstract}
This paper provides the first finite-dimensional characterization for the complete set of full-block, circle criterion multipliers. We consider the interconnection of a discrete-time, linear time-invariant system in feedback with a non-repeated, sector-bounded nonlinearity. Sufficient conditions for stability and performance can be derived using: (i) dissipation inequalities, and  (ii)  Quadratic Constraints (QCs) that bound the input/output pairs of the nonlinearity. Larger classes of QCs (or multipliers) reduce the conservatism of the conditions.  Full-block, circle criterion multipliers define the complete set of all possible QCs for non-repeated, sector-bounded nonlinearities. These provide the least conservative conditions. However, full-block multipliers are defined by an uncountably infinite number of constraints and hence do not lead to computationally tractable solutions if left in this raw form. This paper provides a new finite-dimensional characterization for the set of full-block, circle criterion multipliers.  The key theoretical insight is: the set of all input/output pairs of non-repeated sector-bounded nonlinearities is equal to the set of all incremental pairs for an appropriately constructed piecewise linear function.  Our new description for the complete set of multipliers only requires a finite number of matrix copositivity constraints. These conditions have an exact, computationally tractable implementation for problems where the nonlinearity has small input/output dimensions $(\le 4)$.  
We illustrate the use of our new characterization via a simple example.


\end{abstract}


\section{Introduction}\label{sec:intro}
\label{sec:introduction}
This paper considers systems formed by the interconnection of a known (nominal) discrete-time system and a static, memorlyess nonlinearity.  The analysis objective is to assess the stability and induced gain of this interconnection. Integral Quadratic Constraints  (IQCS)  \cite{megretski2002,veenman2016,hu2017}, can be used to bound the input/output behavior of the nonlinearity.  Sufficient conditions for stability and performance can then be given using dissipation inequalities that incorporate the  IQCs.

We specifically focus on  nonlinearities defined by a non-repeated collection of scalar functions, each of which satisfies a given sector bound. For this class of nonlinearities, there are widely known sets of Quadratic Constraints (QCs) to bound the input/output pairs at each point in time, e.g.
see Section 8.1 of \cite{BEFB94}. The QC
is defined by a matrix, also known as a multiplier. In general, larger sets of multipliers provide less conservative analysis results but with more computational cost.  The largest class of QCs for non-repeated, sector-bounded nonlinearities is given by the full-block
"circle criterion" multiplier  \cite{veenman2014,veenman2016,Fetzer2017IFAC,Fetzer2017RNC,iwasaki1998,scherer2001}. In fact, full block multipliers define the complete set of QCs for non-repeated, sector-bounded nonlinearities. However, this set of QCs is described by an infinite number of conditions and hence it is not computationally tractable.

Our key contribution, given in Section~\ref{sec:main},  is to provide an alternative characterization for the set of full-block multipliers.  We first show that the set of all input/output  pairs from the non-repeated sector-bounded nonlinearities is equal to the set of incremental pairs from an appropriately constructed piecewise linear function (Lemma~\ref{lem:SetEquivalence}).  This equivalence means that any valid QC for one set of pairs is valid for the other set of pairs. Next we derive the complete set of QCs for the incremental pairs of the piecewise linear function (Theorem~\ref{thm:Mcmp}). Combining these results leads to our alternative characterization for the full-block QCs (Corollary~\ref{cor:equal}).

Our alternative characterization is described by $4^m$ copositivity conditions constraints that are $2m\times 2m$ where $m$ is the input/output dimension of the nonlinearitiy. Checking copositivity of a matrix is a co-NP-complete problem~\cite{Murty1987}.  There exist well-known sufficient conditions for copositivity and these relaxations are exact for $m\le 4$ \cite{Berman2003,Diananda1962}.  
Hence our  representation for the complete set allows for an exact computational implementation when $m \le 4$. 
Finally, we provide an example to illustrate these results in Section~\ref{sec:numex}.



\vspace{10pt}
\noindent
\textbf{Notation:}
The sets of real, $n\times1$ vectors and $m\times n$ matrices are denoted by $\R^{n}$ and $\R^{n\times m}$. The sets $\Rn^n$ and $\Rn^{n\times m}$ denote vectors and matrices of the given dimensions with non-negative entries. Entries into vectors and matrices are denoted by subscripts, e.g. $x_i$ denotes the $i$-th entry of a vector $x\in \R^n$.   The set of real symmetric, $n\times n$ matrices is denoted by $\Sym^n$.  If $M\in\Sym^n$ then $M\succeq 0$,  $M\succ 0$, $M\preceq 0$, and $M\prec 0$ denote that $M$ is positive semidefinite, positive definite, negative semidefinite, and negative definite, respectively.  If $x\in \R^n$ then $M:=\diag{x}\in \R^{n\times n}$ is a diagonal matrix with $M_{ii} = x_i$.  More generally, $\diag{ \{\alpha,\beta\}^n }$ denotes the set of all $n\times n$ diagonal matrices with either $\alpha$ or $\beta$ along the diagonal. There are $2^m$ such matrices. Finally,  $\diag{[\alpha,\beta]^n}$ denotes the set of all $n\times n$ diagonal matrices with diagonal entries in the interval $[\alpha,\beta]$.
A real, vector-valued sequence is a function 
$x:\N \to \R^n$ where $\N$ denotes the set of nonnegative integers. The set of all such sequences is denoted by $\ell^n$. The inner product of two 
sequences $v$, $w\in \ell^n$ is defined by $\langle v ,w\rangle :=\sum_{k=0}^\infty v(k)^\top w(k)$. 
The inner product space containing only sequences with $\langle v,v\rangle < \infty$ is denoted $\ell_2^n$. The corresponding norm is $\norm{v}_2:=\sqrt{\langle v,v\rangle}$.


\section{Problem Formulation}\label{sec:problem}
Consider the uncertain system, shown in Fig.~\ref{fig:UncSys}, built from the feedback interconnection of a known (nominal) system $G$ and a perturbation $\Delta_\Phi$. This interconnection is denoted as $F_u(G,\Delta_\Phi)$. The nominal system $G$ is defined by the following discrete time, linear time-invariant dynamics:
\begin{equation}\label{eq:LTINom}
\begin{split}
        x(k+1) &= A \, x(k) \,\, + B_1 \,  w(k) 
         \,\,\, + B_2 \,  u(k)\\
        v(k) &= C_1 \,  x(k) +D_{11}\,  w(k) +D_{12}\,  u(k)\\
        y(k) &= C_2 \,  x(k) + D_{21}\,  w(k) + D_{22}\,  u(k). \\
    \end{split}
\end{equation}
In~\eqref{eq:LTINom}, $x(k)\in\R^{n_x}$
is the state-vector at time $k\in \field{N}$. The vectors $u(k)\in \R^{n_u}$ and $w(k)\in \R^{n_w}$ are the system inputs, while $y(k)\in\R^{n_y}$ and $v(k) \in \R^{n_v}$ are the outputs at time $k\in \field{N}$, respectively. The state space matrices are real-valued with appropriate dimensions, e.g., $A\in \R^{n_x\times n_x}$. The remainder of the paper assumes $n_v =n_w$. To simplify notation, we denote this dimension by $m:=n_v=n_w$. 

\begin{figure}[ht!]
\centering
\input{figures/UncSys}
\caption{Feedback interconnection $F_u(G,\Delta_\Phi)$ of an LTI system $G$ and perturbation $\Delta_\Phi$ defined by a repeated nonlinearity $\Phi$.}
\label{fig:UncSys}
\vspace{0pt}
\end{figure}



The perturbation $\Delta_\Phi: \ell^m\to \ell^m$ is defined by non-repeated scalar nonlinearities. Specifically, let $\phi_i:\R \to \R$ be scalar functions for $i=1,\ldots,m$. Define the non-repeated nonlinearity $\Phi:\R^{m}\to \R^{m}$ by $\Phi(v):= [\phi_1(v_1), \ldots, \phi_m(v_{m})]^\top$.  The perturbation $\Delta_\Phi$ maps the input signal $v\in \ell^m$ to the output $w\in\ell^m$ by application of the non-repeated nonlinearity at each point in time: $w(k) = \Phi\left( v(k) \right)$.
Here, we focus on sector-bounded nonlinearities.

\vspace{10pt}
\begin{defn}[Sector Bound]
    \label{def:sb}
    A scalar function $\phi:\R\rightarrow \R$  is in the sector $[\alpha,\beta]$, for some $\alpha \le \beta$, if 
    \begin{equation}\label{eq:sb}
        (\phi(x) - \alpha x)\, (\beta x -\phi(x))\ge 0\quad \forall x\in \R.
    \end{equation}
    If a collection of scalar functions $\{ \phi_i \}_{i=1}^m$ each satisfy \eqref{eq:sb} then we also say that the corresponding non-repeated nonlinearity $\Phi:\R^{m}\to\R^{m}$ is in the sector $[\alpha,\beta]$, denoted $\Phi \in \text{sec}[\alpha, \beta]^{m}$.

\end{defn}
\vspace{10pt}

We will analyze the the properties of the interconnection $F_u(G, \Delta_\Phi)$ with sector-bounded perturbations.  In particular, we will consider well-posedness, internal stability and finite induced gain. These three concepts are defined next.


\vspace{10pt}
\begin{defn}[Well-Posedness]\label{def:wp}
    The feedback interconnection $F_u(G,\Delta_\Phi)$ is \emph{well-posed} if for all initial conditions $x(0)\in \R^{n_x}$ and inputs $u\in \ell^{n_u}$, there exists unique $x\in \ell^{n_x}$, $v\in \ell^{m}$, $w\in \ell^{m}$, and $y\in \ell^{n_y}$ that satisfy \eqref{eq:LTINom} and $w =\Delta_\Phi(v)$. Moreover, the signals $(x,v,w,y)$ must have causal dependence on $u$.
\end{defn}
\vspace{10pt}
\begin{defn}[Internal Stability]\label{def:internalstab}
A well-posed interconnection $F_u(G,\Delta_\Phi)$ is \emph{internally stable} if $x(k)\to 0$ as $k\to \infty$ from any initial condition $x(0)\in\R^{n_x}$ with $u(k)=0$ for $k\in \N$. In other words, $F_u(G,\Delta_\Phi)$ is internally stable if $x=0$ is a globally asymptotically stable equilibrium point with no external input.
\end{defn}
\vspace{10pt}
\begin{defn}[Induced $\ell_2$ Gain]\label{def:inducedgain}
A well-posed interconnection $F_u(G,\Delta_\Phi)$ has \emph{finite induced-$\ell_2$ gain} if there exists  $\gamma<\infty$ such that the output $y$ generated by any $u\in \ell_2^{n_u}$ with $x(0)=0$ satisfies $\|y\|_2 \le \gamma \, \|u\|_2$.  We denote the infimum of all such $\gamma$ by $\|F_u(G,\Delta_\Phi)\|_{2 \to 2}$.
\end{defn}
\vspace{10pt}


Note that the interconnection $F_u(G,\Delta_\Phi)$ involves an implicit equation.  Specifically, the second equation in \eqref{eq:LTINom} combined with $w(k)=\Phi\left( v(k) \right)$ yields:
\begin{align}
   \label{eq:WellPosed}
   v(k) = C_{1}\,x(k) 
   +D_{11}\, \Phi\left( v(k) \right)
   + D_{12} \,u(k).
\end{align}
The dynamic system $F_u(G,\Delta_\Phi)$ is well-posed as stated in Definition~\ref{def:wp} if there exists a unique solution $v(k)$ to  \eqref{eq:WellPosed}
for all values of $x(k)$ and $u(k)$.

\section{Background}\label{sec:back}

This section first defines quadratic constraints (QCs) for bounding the input/output pairs of nonlinear functions.
We will use these QCs and dissipation inequalities to provide a sufficient condition for internal stability and finite induced $\ell_2$-gain. Finally, we will review existing sets of QCs for non-repeated, sector-bounded nonlinearities. We also comment on the case of repeated, sector-bounded nonlinearities.

\subsection{Quadratic Constraints}
\label{ss:qc}

Quadratic constraints can be used to outer bound a set $\Se\subset \R^{2m}$. This fact is formally defined next.
\vspace{10pt}
\begin{defn}\label{def:QCset}
    Let $M \in \Sym^{2m}$ be given. The set $\Se \subset \R^{2m}$ satisfies the quadratic constraint defined by $M$ if the following inequality  holds for all
    $z \in \Se$:
    \begin{equation}\label{eq:qc}
        z^\top \, M z \ge 0
    \end{equation}
    The matrix $M$ is called the multiplier associated with the QC.
\end{defn}
\vspace{10pt}

We let $\QC{M}$ denote the set of all vectors $z \in \R^{2m}$ that satisfy the QC in \eqref{eq:qc} defined by $M\in\Sym^{2m}$. Then the set $\Se$ satisfies the QC defined by $M$ if $\Se \subset \QC{M}$. In other words, the QC defines an outer approximation (a superset) of all possible vectors in $\Se$.   




We can use Definition~\ref{def:QCset} to bound the input/output behavior of a nonlinear function. To make this precise, define the graph of a function $\Phi:\R^{m} \to \R^{m}$ by the following set of input/output pairs:
\begin{align}
    \Se(\Phi) := \left\{ 
    \bmtx v\\ w\emtx \in \R^{2m}
    \, : \,
    v \in \R^{m}, \, w = \Phi(v)
    \right\}.
\end{align}
Then $M\in \Sym^{2m}$ defines a valid QC for the nonlinearity $\Phi$, i.e., $\Se(\Phi) \subset \QC{M}$, if
\begin{align}
\bmtx v \\ w \emtx^\top M 
\bmtx v \\ w \emtx \ge 0  
\quad\quad 
\forall v \in \R^m, 
\,\,
w = \Phi(v)
\end{align}

We will focus on  QCs for the set of  input/output pairs arising from all non-repeated nonlinearities in the sector $[\alpha,\beta]$. This set is defined as follows:
\begin{align}
\label{eq:Gsec}
    \Se({\mbox{sec}[\alpha,\beta]^m}) := 
    \bigcup_{\Phi \in \mbox{sec}[\alpha,\beta]^{m}} \Se(\Phi)
\end{align}

\subsection{Stability Condition}
\label{ss:stab}

We next state a standard sufficient condition for stability and performance of $F_u(G,\Delta_\Phi)$. The theorem below is for the specific case of non-repeated, sector-bounded nonlinearities. The condition is stated using an affine function $L: \Sym^{n_x} \times \Sym^{2m} \times \R_{\ge 0} \to \Sym^{n_x+m+n_u}$ defined as follows:
\begin{align}
\nonumber
& L(P,M,\gamma^2) := 
\bmtx A^\top P A-P  & A^\top P B_1 
  &  A^\top P B_2 \\ 
  B_1^\top P A & B_1^\top PB_1 
  & B_1^\top P B_2 \\
  B_2^\top P A & B_2^\top P B_1 
& B_2^\top P B_2-\gamma^2 I\emtx \\
\label{eq:LMI}
& \hspace{0.2in}
+ \bmtx C_2^\top \\ D_{21}^\top \\ D_{22}^\top\emtx
  \bmtx C_2^\top \\ D_{21}^\top \\ D_{22}^\top \emtx^\top
+  \bmtx C_1^\top & 0 \\ D_{11}^\top & I  \\ D_{12}^\top & 0 \emtx
M
  \bmtx C_1^\top & 0 \\ D_{11}^\top & I  \\ D_{12}^\top & 0 \emtx^\top.
\end{align}

\vspace{0.1in}
\begin{mytheo}    
    \label{thm:StabPerfConf}
    Consider the interconnection $F_u(G,\Delta_\Phi)$ 
    with the LTI system $G$ defined in \eqref{eq:LTINom}.
    Assume the following:
    \begin{enumerate} 
        \item $D_{11} = 0$.
        \item  All non-repeated nonlinearities in the sector $[\alpha,\beta]$ satisfy the QC defined by $M\in \Sym^{2m}$, i.e., $\Se({\mbox{sec}[\alpha,\beta]^m}) \subset \QC{M}$.
        \item There exists a matrix $P \in \Sym^{n_x}$ and
        scalar $\gamma>0$ such that $P\succeq 0$ and $L(P,M,\gamma^2) \prec 0$.  
    \end{enumerate}
    Given these assumptions, if $\Phi: \R^{m} \to \R^{m}$ is any non-repeated nonlinearity in the sector $[\alpha,\beta]$ then
    $F_u(G,\Phi)$ is well-posed and internally stable with 
    $\|F_u(G,\Delta_\Phi)\|_{2\to 2} < \gamma$.
\end{mytheo}    
\begin{proof}
We have $D_{11}=0$ by Assumption 1, so there exists a unique solution $v(k)$ to  \eqref{eq:WellPosed}
for all values of $x(k)$ and $u(k)$. This implies that the interconnection $F_u(G,\Delta_\Phi)$ is well-posed for any $\Phi \in \text{sec}[\alpha,\beta]^m$.  Hence, the interconnection has a unique  solution $(x, w, v, y)$ for any given initial condition $x(0)\in\R^{n_x}$ and input $u \in \ell_2^{n_u}$.    

The remainder of the proof relies on standard Lyapunov and dissipation inequality arguments. A similar result and proof are given in Theorem 6 of \cite{Noori2024}. A sketch for the induced-gain bound is given for completeness. Define a storage function by $V\left(x\right) := x^\top P x$. Multiply the LMI $L(P,M,\gamma^2)\preceq 0$
on the left and right
by $[x(k)^\top, \, w(k)^\top, \, u(k)^\top]$ and its transpose, respectively.  Applying the  dynamics~\eqref{eq:LTINom} gives the following dissipation inequality:
\begin{align*}
   V\left(x(k+1)\right)  
   & - V\left(x(k)\right)  +  y(k)^\top y(k) \\
   & + \bmtx v(k) \\ w(k) \emtx^\top
    M \bmtx v(k) \\ w(k) \emtx 
    \le \gamma^2\,  u(k)^\top u(k).
\end{align*}
The last term on the left side is non-negative due to Assumption 2. Summing the dissipation inequality from $k=0$ to $k=T-1$ yields:
\begin{align*}
   V\left(x(T)\right) - V(x(0)) +  \sum_{k=0}^{T-1} y(k)^\top y(k)  \le \gamma^2 \sum_{k=0}^{T-1}  u(k) ^\top u(k). 
\end{align*}
Note that $V(x(T))\ge 0$ because $P \succeq 0$. Moreover, the right side is upper bounded by $\gamma^2 \|u\|_2^2$. This yields the following bound for all $T > 0$.
\begin{align*}
   \sum_{k=0}^{T-1} y(k)^\top y(k)  
   \le \gamma^2 \|u\|_2^2 + V(x(0)).
\end{align*}
This  bound implies that the interconnection has a finite induced-$\ell_2$ gain: $\|F_u(G,\Delta_\Phi)\|_{2\to 2} \le \gamma$.  A few more details are required to show the strict inequality $\|F_u(G,\Delta_\Phi)\|_{2\to 2} < \gamma$
as stated in the theorem.
\end{proof}
\vspace{0.1in}

We assumed $D_{11}=0$ in     Theorem~\ref{thm:StabPerfConf}
for simplicity to ensure well-posedness.
This assumption can be relaxed under additional assumptions on the nonlinearity. For example, if $\Phi$ is locally Lipschitz then there are simple sufficient conditions that ensure well-posedness. See Lemma 1 in \cite{Richardson2023}, which relies on results in \cite{valmorbida18,zaccarian02}.  

\subsection{QCs for Sector-Bounded Nonlinearities}
\label{ss:qc4sb}



This subsection reviews existing sets of QCs for non-repeated sector-bounded nonlinearities. First, 
we present a standard set of QCs that follows directly from the sector-bound inequality~\eqref{eq:sb}. See Section 8.1 of \cite{BEFB94} for details.

\vspace{10Pt}
\begin{lem}\label{lem:Md}
Define the following subset of $\Sym^{2m}$:
\begin{align}\label{eq:Mab}
\begin{split}    
    \Md := \bigg\{ 
      M \in \Sym^{2m} \, : \, 
    & M=\bmtx -2\alpha\beta \, \Lambda & 
    (\alpha+\beta) \, \Lambda  \\ 
    (\alpha+\beta) \, \Lambda & -2\Lambda \emtx,
     \\ 
     & \Lambda = \text{diag}(\lambda), \, \lambda \in \R^{m}, \lambda_i \ge 0 \bigg\}.
\end{split}
\end{align}
Any $M \in \Md$ defines a valid QC for all non-repeated nonlinearities in the sector $[\alpha,\beta]$. In other words,
if $M \in \Md$ then $\Se({\mbox{sec}[\alpha,\beta]^m}) \subset \QC{M}$. 
\end{lem}
\begin{proof}
    Consider any $M \in \Md$. If $\Phi$ is a non-repeated nonlinearity in  $\text{sec}[\alpha, \beta]^{m}$, then any input/output pair $v \in \R^{m}$ and $w=\Phi(v)$ satisfy:
    \begin{equation}\label{eq:sumsec}
         \bmtx v\\ w \emtx^\top M
         \bmtx v\\ w \emtx = 
        2\sum_{i=1}^{m} \lambda_{i}
        (w_i-\alpha v_i) \, (\beta v_i - w_i) \ge 0.
    \end{equation}
    The sum is nonnegative because $\lambda_{i} \ge 0$ and $v_i$, $w_i=\phi_i(v_i)$ satisfy the sector-bounded inequality \eqref{eq:sb}. 
\end{proof}
\vspace{10pt}

This set of multipliers is often used since it yields computationally tractable conditions when combined with Theorem~\ref{thm:StabPerfConf}. However, this class of QCs gives conservative bounds for non-repeated, sector-bounded nonlinearities.  Full-block multipliers \cite{veenman2014,veenman2016,Fetzer2017IFAC,Fetzer2017RNC,iwasaki1998,scherer2001}  provide tighter bounds and thus less conservative stability/performance conditions. The next lemma reviews the full-block "circle criterion" multipliers for non-repeated sector-bounded nonlinearities.  Details can be found in  Section 4.1 of \cite{Fetzer2017RNC} or Section 5.8.2 of \cite{veenman2016}. 

\vspace{10pt}
\begin{lem}\label{lem:Mfb}
Define the following subset of $\Sym^{2m}$:
\begin{equation}\label{eq:Mfb}
\begin{split}    
    \Mfb := \Bigg\{  M\in \Sym^{2m}:\bmtx I_{m} \\ \Gamma \emtx^\top M \bmtx I_{m} \\ \Gamma \emtx \succeq 0  \\\forall\, \Gamma \in \text{diag}([\alpha, \beta]^{m}) \Bigg\}.   
    \end{split}
\end{equation}    
$M$ defines a valid QC for all non-repeated nonlinearities in the sector $[\alpha,\beta]$ if and only if $M \in \Mfb$.
\end{lem}
\begin{proof}

($\Leftarrow$) Consider any $M \in \Mfb$. Let $v \in \R^{m}$ and $w=\Phi(v)$ be any input/output pair from a nonlinearity
$\Phi\in \text{sec}[\alpha, \beta]^{m}$. Define a diagonal matrix $\Gamma \in \R^{m\times m}$ by:
\begin{align}
    \Gamma_{ii} := \left\{
        \begin{array}{cc}
            \frac{w_i}{v_i}  & \mbox{if } v_i \ne 0 \\
            \frac{1}{2} (\alpha+\beta) & \mbox{otherwise}
        \end{array}
    \right. .
\end{align}
It follows from this definition and the sector-bound \eqref{eq:sb} that $w = \Gamma v$, and $\Gamma \in \text{diag}([\alpha, \beta]^{m})$. 
Using this construction, the pair $(v,w)$ satisfies:
\begin{align}
    \label{eq:fbrelation}
  \bmtx v\\ w \emtx^\top M \bmtx v\\ w \emtx = 
   v^\top  \bmtx I_{m} \\ \Gamma \emtx^\top M \bmtx I_{m} \\ \Gamma \emtx   v \ge 0.
\end{align}
The result is non-negative because
$M \in \Mfb$ and $\Gamma \in \text{diag}([\alpha, \beta]^{m})$. Hence the QC defined by any $M\in\Mfb$ is valid for input/output pairs from all non-repeated nonlinearities in 
$\text{sec}[\alpha, \beta]^{m}$.

\vspace{10pt}
($\Rightarrow$) Consider any symmetric $M \notin \Mfb$. Then there exists
$\bar \Gamma \in \text{diag}([\alpha, \beta]^{m})$ and $\bar v \in 
\R^m$ such that:
\begin{align}
  \label{eq:fbcounterex}
   \bar{v}^\top  \bmtx I_{m} \\ \bar\Gamma \emtx^\top 
   M \bmtx I_{m} \\ \bar\Gamma \emtx \bar{v} < 0. 
\end{align}
Let $\bar{w}:=\bar{\Gamma} \bar{v}$ and define scalar functions $\bar{\phi}_i:\R \to \R$ by $\bar{\phi}_i(v) = \bar{\Gamma}_{ii} \, v$. The corresponding non-repeated nonlinearity $\bar{\Phi}:\R^m \to \R^m$ is, by construction, in $\text{sec}[\alpha, \beta]^{m}$ and satisfies $\bar{w} = \bar{\Phi}(\bar{v})$. 
By \eqref{eq:fbcounterex}, this function fails to satisfy the QC defined by $M$ at the input/output pair $(\bar{v},\bar{w})$.

\end{proof}
\vspace{10pt}


Lemma~\ref{lem:Mfb} states that $\Mfb$ provides the complete set of valid QCs for the set of non-repeated, sector-bounded nonlinearities. Note that $\Mfb$ is defined by an infinite number of constraints on $M$. Specifically, $[\alpha, \beta]^{m}$ is an $m$-dimensional hypercube and $\Mfb$ has one constraint for each $\Gamma \in \text{diag}([\alpha, \beta]^{m})$. Several subsets of $\Mfb$ have been proposed to yield  computationally tractable analysis conditions. Here, we review one specific convex relaxation of $\Mfb$ (see Section 4.1 of \cite{Fetzer2017RNC}). This relaxation  enforces constraints on $M$ using the finite set $\Gamma \in \text{diag}(\{\alpha, \beta\}^{m})$, i.e., using diagonal matrices defined on the $2^m$ vertices of the hypercube $[\alpha, \beta]^{m}$.  This relaxation, given in the next lemma, requires an additional sign definiteness condition on the lower right block of $M$.

\vspace{10pt}
\begin{lem}\label{lem:Mfc}
    Define the following subset of $\Sym^{2m}$:
    \begin{equation}\label{eq:Mfc}
    \begin{split}
    \Mfc := \Bigg\{ M \in \Sym^{2m}: M=\bmtx Q&S\\S^\top & R\emtx, \, R\in \Sym^m, 
    R \prec 0,\\ 
    \bmtx I_{m} \\ \Gamma \emtx^\top M \bmtx I_{m} \\ \Gamma \emtx \succeq 0
    \,, \forall \Gamma \in \diag{\{\alpha, \beta\}^m} \Bigg\}.
    \end{split}
\end{equation}
Any $M \in \Mfc$ defines a valid QC for all non-repeated nonlinearities in the sector $[\alpha,\beta]$. 
In other words, if $M \in \Mfc$ then $\Se({\mbox{sec}[\alpha,\beta]^m}) \subset \QC{M}$. 
\end{lem}
\begin{proof}
    To simplify notation, define the function $h_M:[\alpha,\beta]^m \to \Sym^{2m}$ by
    \begin{align}
        h_M(x) := \bmtx I_m \\ \diag{x} \emtx^\top M \bmtx I_m \\ \diag{x} \emtx.
    \end{align}    
    The set $\Mfc$ only enforces $h_M(x) \succeq 0$ on the vertices
    $\{\alpha, \beta\}^m$. We will show that this, combined with $R \prec 0$, is sufficient to ensure $h_M(x) \succeq 0$ for all points
    in the hypercube $\{\alpha, \beta\}^m$.  This will imply that $\Mfc \subset \Mfb$ and hence any $M\in \Mfc$ defines a valid QC for non-repeated nonlinearities in the sector $[\alpha,\beta]$ by Lemma~\ref{lem:Mfb}.

    First, we show that if $M\in \Mfc$ then $h_M$ is a concave function. Consider any $x$, $y\in \diag{[\alpha,\beta]^m}$ and any $\theta\in [0,1]$.    
    Let $\Gamma_x:=\diag{x}$  and $\Gamma_y:=\diag{y}$ 
    denote the corresponding diagonal matrices.  The following relation can be shown by re-grouping terms:
    \begin{align}
    \begin{split}        
        & h_M\left(\theta x + (1-\theta)y \right)
        - \theta h_M(x) 
        - (1-\theta) h_M(y) \\
        & \quad \quad
        = -\theta (1-\theta) \, (\Gamma_x - \Gamma_y) R (\Gamma_x - \Gamma_y).
    \end{split}
    \end{align}
    The right side is positive semidefinite because $R\preceq 0$ and $\theta \in [0,1]$. Hence we have that $h_M$ is concave:
    \begin{align}
         h_M\left(\theta x + (1-\theta)y \right) \succeq
                \theta h_M(x) + (1-\theta) h_M(y).
    \end{align}
    
    Next, let $\{\bar{x}^{(i)}\}_{i=1}^{2^m}$ denote 
    the vectors in the set $\{\alpha,\beta\}^m$.
    Recall that a hypercube  is equal to the convex hull of its vertices (Section 2.2.4 of \cite{boyd2004}). Hence for any $x \in [\alpha,\beta]^m$ there exists $\{ \theta_i \}_{i=1}^{2^m}$    such that $\theta_i \ge 0$, $\sum_{i=1}^{2^m} \theta_i = 1$ and $x = \sum_{i=1}^{2^m} \theta_i \bar{x}^{(i)}$.
    The concavity of $h_M$ implies:
    \begin{align}
        h_M(x) \succeq \sum_{i=1}^{2^m} \theta_i h_M( \bar{x}^{(i)} )
    \end{align}
    The right side is positive semidefinite because $M\in \Mfc$ implies
    $h_M(\bar{x}^{(i)}) \succeq 0$ for $i=1,\ldots,2^m$. It follows that $h_M(x) \succeq 0$ for any  $x\in \diag{ [\alpha,\beta]^m}$ and hence $M\in \Mfb$.
\end{proof}
\vspace{10pt}

It can be easily verified that $\Md \subset \Mfc$. Hence the convex relaxation $\Mfc$ will provide, in general, less conservative bounds than the standard set $\Md$. Other subsets of $\Mfb$ are given in \cite{Fetzer2017RNC} and \cite{veenman2016} using a generalization of Polya's theorem. These provide less conservative bounds than $\Mfc$ although with additional computational cost.

\subsection{Comment on Repeated Nonlinearities}\label{sec:repnl}

We briefly comment on the case of repeated nonlinear functions. These are functions $\Phi:\R^{m}\to \R^{m}$ of the form $\Phi(v):= [\phi(v_1), \ldots, \phi(v_{m})]^\top$
for some $\phi:\R\to \R$.  We denote the set of all \emph{repeated} nonlinearities in the sector $[\alpha,\beta]$ by $\text{repsec}[\alpha,\beta]^m$.  

Repeated nonlinearities are special cases of non-repeated nonlinearities. Hence  $\text{repsec}[\alpha,\beta]^m \subset \text{sec}[\alpha,\beta]^m$ and
$\Se({\mbox{repsec}[\alpha,\beta]^m})
 \subset \Se({\mbox{sec}[\alpha,\beta]^m})$.
This implies that every QC for the set of non-repeated nonlinearities is also a valid QC for all repeated nonlinerities. One might think that there are additional valid QCs for repeated nonlinearities (since they define a smaller set of input/output pairs).  However, the next lemma states that the set of full block multipliers $\Mfb$ defined in \eqref{eq:Mfb} is also the complete class for $\text{repsec}[\alpha,\beta]^m$.
\vspace{10pt}
\begin{lem}\label{lem:Mfbnr}
A matrix $M\in\Sym^{2m}$ defines a valid QC for all repeated nonlinearities in the sector $[\alpha,\beta]$ if and only if $M \in \Mfb$. Hence $\Mfb$ defines the complete set of QCs for both repeated and non-repeated nonlinearities in the sector $[\alpha,\beta]$.
\end{lem}
\begin{proof}

($\Leftarrow$) This direction of the proof is the same as given for Lemma~\ref{lem:Mfb}.

\vspace{10pt}
($\Rightarrow$) Consider any $M \notin \Mfb$. Then there exists
$\bar \Gamma \in \text{diag}([\alpha, \beta]^{m})$ and $\bar v \in 
\R^m$ such that:
\begin{align}
  \label{eq:repfbcounterex}
   \bar{v}^\top  \bmtx I_{m} \\ \bar\Gamma \emtx^\top 
   M \bmtx I_{m} \\ \bar\Gamma \emtx \bar{v} < 0. 
\end{align}
The inequality is strict.  Thus, it is possible to perturb the entries of $\bar{v}$, if needed, to ensure that all the entries are unique. Next define $\bar{w}:=\bar{\Gamma} \bar{v}$ and note that the pairs $(\bar{v}_i,\bar{w}_i)$ lie in the sector $[\alpha,\beta]$ for $i=1,\ldots, m$.  Finally, define a scalar function $\bar{\phi}:\R \to \R$ to linearly interpolate the points $\{ (\bar{v}_i,\bar{w}_i)\}_{i=1}^m$ and $(0,0)$. The corresponding repeated nonlinearity $\bar{\Phi}:\R^m \to \R^m$ is, by construction, in $\text{repsec}[\alpha, \beta]^{m}$ and satisfies $\bar{w} = \bar{\Phi}(\bar{v})$.
By \eqref{eq:repfbcounterex}, the function $\bar{\Phi}$ fails to satisfy the QC defined by $M$ at the input/output pair $(\bar{v},\bar{w})$.
\end{proof}
\vspace{10pt}

The set of repeated nonlinearities is a strict subset of the set of non-repeated nonlinearities.
However, any input/output pair from a non-repeated sector-bounded function can be perturbed by an arbitrarily small amount to generate an input/output pair form a repeated, sector bounded function.  In other words, the closure of 
$\Se({\mbox{repsec}[\alpha,\beta]^m})$ is equal
to $\Se({\mbox{sec}[\alpha,\beta]^m})$. This explains why the two sets of nonlinearities have the same complete class of valid QCs.

\section{Main Result}\label{sec:main}
This section derives a new description of full-block multipliers for non-repeated nonlinearities in the sector $[\alpha,\beta]$. The new description is defined using the incremental graph of a specific piecewise linear function. This leads to a desription given in terms of a finite-number of copositivity conditions.

\subsection{Incremental Graph for a Piecewise Linear Function}


For a given $\alpha \le \beta$, define a piecewise-linear function $f_{\alpha\beta}:\R \rightarrow \R$ by:
\begin{equation}\label{eq:fab}
    f_{\alpha \beta}(x):= 
    \begin{cases}
    \alpha x \quad x\le 0 \\
    \beta x \quad x>0
    \end{cases}. 
\end{equation}
The corresponding repeated, piecewise-linear function $F_{\alpha\beta}: \R^m\rightarrow \R^m$ is defined by applying $f_{\alpha\beta}$ to the elements of $v$.
The incremental graph of $F_{\alpha\beta}$ is defined by the set of all its incremental input/output pairs
\begin{equation}\label{eq:inc}
\begin{split}
     \Inc(F_{\alpha\beta}):= \Bigg\{& 
     \bmtx \bar{v} - \hat{v} \\
     \bar{w}-\hat{w}\emtx\in \R^{2m} 
     \, :  
     \exists\, \bar{v}, \, \hat{v} \in \R^m 
    \mbox{ s.t.} \\
     & \bar{w} = F_{\alpha\beta}(\bar{v})
     \mbox{ and }
     \hat{w} = F_{\alpha\beta}(\hat{v})
     \Bigg\}.
\end{split}
\end{equation}


The piecewise-linearity of $f_{\alpha\beta}$~\eqref{eq:fab} implies that all incremental input/output pairs are confined to the sector $[\alpha,\beta]$. This is illustrated in Fig.~\ref{fig:Viz} for $\alpha=1$ and $\beta =3$. The left  subplot shows $f_{\alpha\beta}$ along with the points $(\bar{v},\bar{w}) = (1.5,4.5)$ and $(\hat{v},\hat{w}) = (-2,-2)$. The right plot shows the sector bound (shaded blue) for this choice of $[\alpha,\beta]$.  The incremental pair $(\bar{v}-\hat{v}, \bar{w}-\hat{w}) = (3.5,6.5)$ is also shown (red dot). This pair lies in the sector bound because the slope is $\frac{6.5}{3.5} \approx 1.86 \in [\alpha,\beta]$. Switching the points, i.e. 
$(\bar{v},\bar{w}) = (-2,-2)$ and $(\hat{v},\hat{w}) = (1.5,4.5)$ flips the sign of the incremental pair to $(\bar{v}-\hat{v}, \bar{w}-\hat{w}) = (-3.5,-6.5)$. In this case, the point lies in the sector in the third quadrant.  

In general, we can select pairs of points on $f_{\alpha\beta}$ to achieve any slope in $[\alpha,\beta]$.  This is the key intuition for the following fact: the set of all incremental input/output pairs of $F_{\alpha\beta}$ is the same as the set of all input/output pairs of non-repeated nonlinearities in $\text{sec}[\alpha,\beta]^m$.  We formalize this statement in the following lemma.


\begin{figure}[h!]
 \centering
 \input{figures/LinePlots}
    \caption{
    The left plot shows $f_{\alpha\beta}(v)$ vs. $v$ for
    $\alpha=1$ and $\beta=3$. The right plot shows the sector bound (shaded blue) for this choice of $[\alpha,\beta]$.     The points $(\bar{v},\bar{w}) = (1.5,4.5)$ and $(\hat{v},\hat{w}) = (-2,-2)$ are also shown on the left along with the increment on the right,  $(\bar{v}-\hat{v}, \bar{w}-\hat{w}) = (3.5,6.5)$ (red dots). } 
\label{fig:Viz}
\end{figure}


\vspace{10pt}
\begin{lem}\label{lem:SetEquivalence}
    The set of all incremental input/output pairs 
    of $F_{\alpha\beta}$ is equal to the set of all input/output pairs from non-repeated nonlinearities in the sector $[\alpha,\beta]$, i.e., $\Inc(F_{\alpha\beta})=\Se(\text{sec}[\alpha, \beta]^m)$.
\end{lem}
\begin{proof}

\noindent
$\Inc(F_{\alpha\beta}) \subseteq \Se(\text{sec}[\alpha,\beta]^m)$: Take any $\bar{v}$, $\hat{v} \in \R^m$ and let $\bar{w} = F_{\alpha\beta}(\bar{v})$ and $\hat{w} = F_{\alpha\beta}(\hat{v})$. Define the increments $dv := \bar{v} - \hat{v}$ and $dw:=\bar{w} - \hat{w}$.  
We will first show that there exists $\gamma_i \in [\alpha,\beta]$ such that $dw_i = \gamma_i \, dv_i$ for $i=1,\ldots,m$.

If both $\bar{v}_i\le 0$ and $\hat{v}_i \le 0$ then $\bar{w}_i = \alpha \bar{v}_i$ and $\hat{w}_i = \alpha \hat{v}_i$. In this case, the increments satisfy $dw_i = \gamma_i \, dv_i$ with $\gamma_i=\alpha$. Similarly, if  both $\bar{v}_i>0$ and $\hat{v}_i>0$ then the increments satisfy $dw_i = \gamma_i \, dv_i$ with $\gamma_i = \beta$. Finally, consider the case where $\hat{v}_i < 0 < \bar{v}_i$. Then $\bar{v}_i = \beta \bar{v}_i$ and $\hat{w}_i = \alpha \hat{v}_i$. The increments for this case satisfy $dw_i = \gamma_i \, dv_i$ where $\gamma_i:= \frac{\beta \bar{v}_i - \alpha \hat{v}_i}{dv_i}$.  The conditions $\hat{v}_i<0<\bar{v}_i$ and $\alpha \le \beta$ imply that $\gamma_i \in [\alpha,\beta]$.
The remaining case $\hat{v}_i > 0 > \bar{v}_i$ is similar.

Next define scalar functions $\phi_i:\R \to \R$ by $\phi_i(v_i) = \gamma_i \, v_i$. The corresponding non-repeated nonlinearity $\Phi:\R^m \to \R^m$ is in $\text{sec}[\alpha, \beta]^{m}$ and satisfies $dw = \Phi(dv)$ by construction.  Hence $\bmtx dv^\top, \, dw^\top\emtx^\top 
\in \Se(\text{sec}[\alpha,\beta]^m)$.

\vspace{10pt}
\noindent
$\Se(\text{sec}[\alpha,\beta]^m) \subseteq \Inc(F_{\alpha\beta})$:  Let $v \in \R^{m}$ and $w=\Phi(v)$ be any input/output pair from a nonlinearity
$\Phi\in \text{sec}[\alpha, \beta]^{m}$. 
We will use the intuition from Figure~\ref{fig:Viz} to construct input/output pairs $(\bar{v},\bar{w})$ and $(\hat{v},\hat{w})$ from $F_{\alpha\beta}$
such that $v=\bar{v}-\hat{v}$ and $w=\bar{w}-\hat{w}$. 
The proof is trivial if $\alpha = \beta$.
\footnote{If $\alpha = \beta$ then simply select $\bar{v}_i := 1+v_i$, $\hat{v}_i:=1$, $\bar{w}_i := \alpha \bar{v}_i$
and $\hat{w}_i := \alpha \bar{w}_i$ for $i=1,\ldots,m$.  Hence $\bar{w}_i = f_{\alpha\beta}(\bar{v}_i)$ and $\hat{w}_i = f_{\alpha\beta}(\hat{v}_i)$. Moreover, 
the corresponding increments
satisfy $dv_i = v_i$ and $dw_i = \alpha v_i = w_i$. We conclude $(v, w)\in \Inc(F_{\alpha\beta})$ for this trivial case. }  Hence we will assume $\alpha<\beta$ in the remainder of the proof.

Define scalars $\gamma_i \in \R$ for $i=1,\ldots,m$ by:
\begin{align}
    \gamma_i := \left\{
        \begin{array}{cc}
            \frac{w_i}{v_i}  & \mbox{if } v_i \ne 0 \\
            \frac{1}{2} (\alpha+\beta) & \mbox{otherwise}
        \end{array}
    \right. .
\end{align}
Note that  $w_i = \gamma_i v_i$ and each $\gamma_i \in [\alpha,\beta]$ by the sector-bound \eqref{eq:sb}. Define the inputs to the piecewise linear function as:
\begin{align}
\bar{v}_i :=\frac{\gamma_i-\alpha}{\beta-\alpha} \, v_i 
\mbox{ and }
\hat{v}_i =\frac{\gamma_i-\beta}{\beta-\alpha} \, v_i
\end{align}
If $v_i>0$ then $\hat{v}_i < 0 < \bar{v}_i$.
In this case, define $\bar{w}_i := \beta \bar{v}_i$ and $\hat{w}_i := \alpha \hat{v}_i$. Hence $\bar{w}_i = f_{\alpha\beta}(\bar{v}_i)$ and $\hat{w}_i = f_{\alpha\beta}(\hat{v}_i)$.
It can also be shown, with some algebra, that the increments
satisfy $dv_i = v_i$ and $dw_i = \gamma_i v_i=w_i$. We conclude $(v, w)\in \Inc(F_{\alpha\beta})$ if $v_i>0$.
If $v_i<0$, then $\hat{v}_i > 0 > \bar{v}_i$. A similar construction follows in this case with $\bar{w}_i := \alpha \bar{v}_i$ and $\hat{w}_i = \beta \hat{v}_i$.

\end{proof}

\subsection{New Characterization for Full-Block Multipliers}

The equality of $\Inc(F_{\alpha\beta})$
and $\Se(\text{sec}[\alpha,\beta]^m)$ (Lemma~\ref{lem:SetEquivalence}) implies that $M$ defines a valid QC for one of these sets if and only if it defines a valid QC for the other. In other words,  $\Inc(F_{\alpha\beta})\subset \QC{M}$ if and only if $\Se(\text{sec}[\alpha,\beta]^m)\subset \QC{M}$.  This follows directly from the definition of a QC (Definition~\ref{def:QCset}). As a consequence, the complete class of QCs for 
$\Inc(F_{\alpha\beta})$ is equal to the complete class $\Se(\text{sec}[\alpha,\beta]^m)$.  We will use this to derive our new characterization for full block multipliers.

First, we derive the complete class of QCs for $\Inc(F_{\alpha\beta})$ by extending the results in~\cite{Noori2024}. This requires some additional notation.  Let $c:=\frac{1}{2}(\alpha+\beta)$ and
$r:=\frac{1}{2}(\beta-\alpha)$ denote the center and radius of the sector $[\alpha,\beta]$. Define the function $g_M:\Sym^m \times \Sym^m \to \Sym^{2m}$ by
\begin{align*}
g_M(\bar{\Gamma},\hat{\Gamma}) =
\bsmtx \bar{\Gamma} & -\hat{\Gamma}\\ 
r \bar{\Gamma} + c I_m& -r\hat{\Gamma} - c I_m\esmtx^\top
M
\bsmtx \bar{\Gamma} & -\hat{\Gamma}\\ 
r \bar{\Gamma} + c I_m
& -r\hat{\Gamma} - c I_m\esmtx.
\end{align*}    
Next, a matrix $M\in \Sym^{2m}$ is said to be copositive if $x^\top M x \ge 0$ for all $x\in \Rn^{2m}$.   Let $\mathcal{COP}^{2m}$ denote the set of all $(2m)\times (2m)$ copositive matrices. Finally, define the following subset of $\Sym^{2m}$:
\begin{align*}
 \Mcmp := \{& M\in \Sym^{2m} \, : \,
 g_M(\bar{\Gamma},\hat{\Gamma}) \in \mathcal{COP}^{2m}\,\\&\forall\,\bar{\Gamma},\hat{\Gamma} \in \diag{\{-1,1\}^m}\},
\end{align*}
Note that each matrix in $\Mcmp$ satisfies a collection of $2^{2m}=4^m$ copositivity conditions. There is one condition for each pair $\bar\Gamma,\hat\Gamma\in \diag{\{-1,1\}^m}$.
Based on these definitions, we can state the complete set of QCs for the incremental pairs
of the piecewise linear function $F_{\alpha\beta}$.
\vspace{10pt}
\begin{mytheo}\label{thm:Mcmp}
    The function $F_{\alpha\beta}$ satisfies an incremental QC defined by $M\in \Sym^{2m}$, i.e., 
    $\Inc(F_{\alpha\beta}) \subset \QC{M}$, if and only if $M\in \Mcmp$.
\end{mytheo}
\begin{proof}

\noindent
($\Leftarrow$) Assume $M\in \Mcmp$. Take any $\bar v, \hat v \in \R^{m}$ and let $\bar w=F_{\alpha\beta}(\bar v)$ and $\hat w = F_{\alpha\beta}(\hat v)$. 
Let $|\bar{v}|$ and $|\hat{v}| \in \R^m_{\ge 0}$ denote the vectors obtained by taking element-wise absolute values of $\bar{v}$ and $\hat{v}$. In addition, define  $\bar{\Gamma}$, $\hat{\Gamma} \in \diag{\{-1,1\}}^m$ by\footnote{ $\text{sign}(\cdot):\R \to \{-1, \, +1\}$ is defined by $\text{sign}(x)=+1$ if $x\ge 0$ and $\text{sign}(x)=-1$ if $x<0$.}
\begin{align}
\bar{\Gamma}_{ii} := \text{sign}(\bar v_i), \quad
\hat{\Gamma}_{ii} := \text{sign}(\hat v_i), \quad
i=1,\ldots,m
\end{align}
Note that $\bar{v} = \bar{\Gamma} |\bar{v}|$
and $\hat{v} = \hat{\Gamma} |\hat{v}|$. Moreover,
we can express $\bar{w}$ and $\hat{w}$ as follows:
\begin{align}
    \label{eq:wbarIncExp}
    \bar{w} & = (r\bar{\Gamma} + c I_m )
     \, |\bar{v}|, \\
    \label{eq:whatIncExp}
    \hat{w} & = (-r\hat{\Gamma} - c I_m ) 
     \, |\hat{v}|, 
\end{align}
where we again use $c:=\frac{1}{2}(\alpha+\beta)$
and $r:=\frac{1}{2}(\beta-\alpha)$. These expressions follow from the piecewise linearity of $f_{\alpha\beta}$.  For example, if $\bar{v}_i<0$ then $\bar{\Gamma}_{ii} = -1$ and
\eqref{eq:wbarIncExp} gives $\bar{w}_i = (c-r)|\bar{v}_i| = -\alpha |\bar{v}_i|$. This is the correct expression for the pair $\bar{v}_i$, $\bar{w}_i = f_{\alpha\beta}(\bar{v}_i)$. The other cases can be checked similarly.

Finally, we can use these expressions to show the following relation for the increments $dv:=\bar{v}-\hat{v}$ and $dw:=\bar{w}-\hat{w}$:
\begin{align}
    \bmtx dv \\ dw \emtx^\top M
    \bmtx dv \\ dw \emtx
    := \bmtx |\bar{v}| \\ | \hat{v} | \emtx^T
       g_M(\bar{\Gamma},\hat{\Gamma})
       \bmtx |\bar{v}| \\ | \hat{v} | \emtx 
       \ge 0
\end{align}
The result is non-negative because
$M \in \Mcmp$ and $\bar{\Gamma}$, $\hat{\Gamma} \in \text{diag}(\{-1,\, +1\}^{m})$. Hence the QC defined by any $M\in\Mcmp$ is valid for all incremental input/output pairs from the repeated piecewise function $F_{\alpha\beta}$.

\vspace{10pt}
\noindent
($\Rightarrow$)  Assume $M\notin \Mcmp$.
Then there exists $\bar{\Gamma}$, $\hat{\Gamma} \in \diag{\{-1,1\}^m}$ and vectors $|\bar{v}|$, 
$|\hat{v}| \in \R_{\ge 0}^m$ such that
\begin{align}
\label{eq:Micounterex}
   \bmtx |\bar{v}| \\ | \hat{v} | \emtx^T
       g_M(\bar{\Gamma},\hat{\Gamma})
       \bmtx |\bar{v}| \\ | \hat{v} | \emtx 
       < 0
\end{align}
The notation for the non-negative vectors is suggestive as we define $\bar{v} := \bar{\Gamma} |\bar{v}|$ and $\hat{v} := \hat{\Gamma} |\hat{v}|$.
We also define $\bar{w}$, $\hat{w} \in \R^m$ by
\begin{align}
    \bar{w} & := (r\bar{\Gamma} + c I_m )
     \, |\bar{v}|, \\
    \label{eq:whatIncExp}
    \hat{w} & := (-r\hat{\Gamma} - c I_m ) 
     \, |\hat{v}|, 
\end{align}
As discussed early in the $(\Leftarrow)$ direction of the proof, the piecewise linearity of $f_{\alpha\beta}$ can be used to show
$\bar{w} = F_{\alpha\beta} (\bar{v})$
and $\hat{w} = F_{\alpha\beta} (\hat{v})$.
Moreover it follows from the definition of 
$g_M$ that \eqref{eq:Micounterex} can be rewritten in terms of the increments $dv:=\bar{v}-\hat{v}$
and $dw:=\bar{w}-\hat{w}$
\begin{align}
    \bmtx dv \\ dw \emtx^\top M
    \bmtx dv \\ dw \emtx <0    
\end{align}
Thus $F_{\alpha\beta}$ has an incremental pair that fails to satisfy the QC defined by $M$.
\end{proof}

\vspace{10pt}

It remains to show that the two sets $\Mcmp$ and $\Mfb$ are identical. This is done in the following Corollary~\ref{cor:equal}.
\vspace{10pt}
\begin{cor}\label{cor:equal}
    The sets $\Mcmp$ and $\Mfb$ are equal for any sector $[\alpha,\beta]$.
\end{cor}
\begin{proof}
Take any $M \in \Sym^{2m}$.  By Theorem~\ref{thm:Mcmp}, $M$ is in $\Mcmp$ if and only if $M$ is a valid  QC for the incremental pairs of $F_{\alpha\beta}$. Moreover, 
Lemma~\ref{lem:SetEquivalence} states
that $\Inc(F_{\alpha\beta})=\Se(\text{sec}[\alpha, \beta]^m)$. Hence  $M$ is in $\Mcmp$ if and only if $M$ defines a valid QC for all input/output pairs from non-repeated nonlinearities in the sector $[\alpha,\beta]$.  Finally, this is equivalent to $M \in \Mfb$ because $\Mfb$ is equal to the complete set of QCs for 
non-repeated nonlinearities in the sector $[\alpha,\beta]$ (Lemma~\ref{lem:Mfb}).
%
%
\end{proof}

\subsection{Computational Characteristics of Multipliers}\label{ss:comp}

The stability and performance condition in Theorem~\ref{thm:StabPerfConf} is a linear matrix inequality in the variables $P\succeq 0$, $\gamma^2$ and any valid parameterization of the multiplier $M\in\M$. Minimizing $\gamma^2$ subject to the LMI constraints yields the following semidefinite program:
\begin{flalign*}
&\qquad \min_{\substack{M,P,\gamma^2}} \gamma^2&\\
&\qquad \mbox{s.t.}&
\end{flalign*}
\vspace{-25pt}
\begin{subequations}
\label{eq:sdp}
\begin{align}
L(P,M,\gamma^2) \prec 0 \label{eq:Lcond}&\,\\ P\succeq 0&\label{eq:Pcond} \\
M\in\mathcal{M}\label{eq:Mcond}&
\end{align}
\end{subequations}
The general setup of~\eqref{eq:sdp} is the same for all previously introduced multiplier sets.  The choice of the multiplier set only impacts the constraint
\eqref{eq:Mcond}.


In general, there is a relation between the complexity of the conditions and the conservativeness of the corresponding stability results.  The following set containments follow from the previous results/discussion:
\begin{align}
    \Md \subset \Mfc \subset \Mfb = \Mcmp
\end{align}
Larger sets of multipliers will provide less conservative bounds, in general, in the SDP \eqref{eq:sdp}.  The set $\Md$ given in \eqref{eq:Mab} only requires
$m$ linear constraints $\Lambda_{ii} \ge 0$,  but this provides the most conservative results. The set $\Mfc$ provides less conservative results. However, it requires the constraint $R\prec 0$ and  $2^m$ additional constraints on $M$ enforced on the vertices of the hypercube $[\alpha,\beta]^m$. Finally, the set $\Mfb$ provides the least conservative result since it contains the complete class of valid multipliers. This set requires an infinite number of constraints on $M$ enforced at each point in the hypercube $[\alpha,\beta]^m$. Hence only approximations of this set can be implemented, e.g. using constraints enforced on a finite gridding of $[\alpha,\beta]^m$.

Our new set $\Mcmp$ is equal to $\Mfb$. Thus it also yields the least conservative results. The important point is that this set is characterized by a finite number of constraints.  Specifically, it requires a total of $2^{2m}=4^m$ copositivity conditions involving $M$. One issue is that checking copositivity of a matrix is a co-NP-complete problem~\cite{Murty1987}. However, any ordinary copositive matrix is the sum of a positive semidefinite matrix and non-negative matrix \cite{Berman2003}. Every copositive matrix of dimension $m \le 4$ is ordinary and the relaxation is exact \cite{Diananda1962}. For dimensions $m\ge 5$, non-ordinary copositive matrices exist, e.g., the Horn-matrix (Example 1.30 in \cite{Berman2003}) and the relaxation is no longer exact.
In summary,  the representation $\Mcmp$ for the complete set allows for an exact computational implementation when $m \le 4$.

\section{Numerical Example}
\label{sec:numex}


This section provides a simple numerical example to demonstrate the application of different sets of full-block multipliers for robust stability and $\ell_2$-gain analysis. 

We consider the interconnection $F_u(G,\Delta_\Phi)$ shown in Fig.~\ref{fig:UncSys}. We assume the perturbation $\Delta_\Phi$ is defined by a $[0,\beta]$ sector-bounded nonlinearity $\Phi: \R^3 \rightarrow \R^3$, i.e., $\Phi\in \text{sec}([0,\beta]^3)$. 
We also assume the nominal part $G$ is a third order, discrete-time system with the following state space matrices:

\begin{align*}
A & :=
\bmtx
0.03 & -0.02 & -0.02\\
0    & -0.06 &  0.02\\
0    &  0.07 &  0.01
\emtx,\\[0.4em]
B_1 & :=
\bmtx
1.21  & 0.95  & 1.38\\
-0.03 & 0.95  & 0.83\\
-0.45 & -0.21 & -0.30
\emtx,
&
B_2 & :=
\bmtx
1.42\\
-0.41\\
-1.08
\emtx,\\[0.4em]
C_1 & :=
\bmtx
-0.66 &  0.48 & 0.42\\
 0.28 & -0.66 & 0.62\\
 0.75 & -0.95 & 0.39
\emtx,
&
C_2 & :=
\bmtx
-1.79 \\ -1.58 \\ -0.92
\emtx^\top,\\[0.4em]
D_{11} & :=
\bmtx
-0.24 & -1.27 & 0.04\\
 0.45 &  0.10 & 0.88\\
 0.50 & -0.93 & -0.02
\emtx,
&
D_{12} & :=
\bmtx
-0.80\\
 0.60\\
 0.61
\emtx,\\[0.4em]
D_{21} & :=
\bmtx
-0.54 & 1.11 & 0.47
\emtx,\,\text{and}
&
D_{22} & := 0.38.
\end{align*}
We will use the LMI condition in Theorem~\ref{thm:StabPerfConf} to assess stability and the induced $\ell_2$ gain of
$F_u(G,\Delta_\Phi)$. Note that Theorem~\ref{thm:StabPerfConf} requires $D_{11}=0$ but this assumption is only required to prove well-posedness. The state-space data given above has $D_{11} \ne 0$ so we will assume the perturbation is such that $F_u(G,\Delta_\Phi)$ is well-posed.  

We bound the behavior of the nonlinearity $\Phi$ with different sets of QCs. Specifically, we consider the multipliers from literature $\Md$ and $\Mfc$ as well as the newly proposed $\Mcmp$. The corresponding semidefinite programs defined by~\eqref{eq:sdp} are solved in Matlab using a combination of \texttt{cvx} \cite{cvx} and \texttt{MOSEK} \cite{mosek} with default settings. Note that each case requires the implementation of the multiplier specific constraint \eqref{eq:Mcond} as described in Section~\ref{ss:comp}.  All calculations were performed on a standard laptop with $4.51\,\text{GHz}$ and $24\,\text{GB}$ memory. 

We first perform the analysis for a nonlinearity with sector bound $\beta = 1$, i.e., $\Phi \in \text{sec}([0, 1]^3)$. This case resembles the incremental stability analysis of a small recurrent neural network with a repeated rectified linear unit (ReLU) activation function, e.g., as in \cite{Revay2021}. 
Analyzing the interconnection using the set of diagonal multipliers $\Md$ yields
an upper bound on the worst-case $\ell_2$-gain of $\gamma_\text{d} = 11.49$ which is calculated in $0.135\,\text{s}$.
Next, we use the convex relaxation $\Mfc$ of the set of full-block multipliers $\Mfb$. This multiplier provides a worst-case gain upper bound $\gamma_\text{c} = 7.844$. The calculation requires a total of $0.163\,\text{s}$.
The last calculation is performed using the new set of QCs defined by $\Mcmp$. We use, in this case, the exact copositivity relaxation described in Section~\ref{ss:comp}.
The new multiplier provides the upper bound $\gamma_\text{inc}= 6.050$. The calculation takes $0.336\,\text{s}$.  Thus the complete set $\Mcmp$ yields an upper bound which is $12.87\%$ less conservative, while requiring approximately twice the computational time of the convex relaxation. Recall that $\Mcmp$ provides the complete class of static QCs for nonlinearities in $\text{sec}[0,1]^m$ and hence it provides the least conservative analysis results when using static QCs.


Next, we analyze the interconnection for nonlinearities in $\text{sec}([0,\beta]^3)$ with $15$ equidistant values of $\beta$ in the interval $0$ to $1.3$. Here, $\beta =0$ corresponds to the nominal system dynamics. The (nominal) gain for $\beta=0$ is $\gamma_\text{nom}= 1.396$. This is calculated with the build-in Matlab function \texttt{norm} using the nominal dynamics $F_u(G,0)$. The remaining upper bounds for $\beta >0$ are calculated as in the first part of the example. Figure~\ref{fig:results} shows the achieved gain bounds over $\beta$. Note that for small values of $\beta$ all gain bounds converge to $\gamma_\text{nom}$ as expected. The gap between $\Md$ and the full multipliers starts to widen for $\beta > 0.6$. The gap between $\Mfc$ and $\Mcmp$ opens notably for $\beta > 0.9$. The largest sector bounds for which $\Md$ and $\Mfc$can find a feasible solution are $\beta_{\Md}=1.17$ and $\beta_{\Md}=1.30$.  These values correspond to the stability margin that can be verified for each class of multipliers. Our complete set $\Mcmp$ is feasible up to $\beta_{\Mcmp} = 1.34$, i.e., we can verify the stability for the largest sector using $\Mcmp$. Thus, we conclude that our proposed set of QC multipliers for sector-bounded nonlinearities $\Mcmp$ provides less conservative stability results than two commonly used sets in literature. Note that more exact approximations of $\Mfb$, as in, e.g.\cite{Fetzer2017RNC}, can provide stability margins and gain bounds which are closer to our complete set $\Mcmp$.



\begin{figure}[h!]
 \centering
 \input{figures/ResultPlot}
    \caption{Achieved gain bound over sector bound $\beta$: $\Md$ (\ref{pl:Md}), $\Mfc$ (\ref{pl:Mfc}), $\Mcmp$ (\ref{pl:Mcmp})}  
\label{fig:results}
\end{figure}



\section{Conclusion}

This paper considered Quadratic Constraints (QCs) for non-repeated and repeated sector-bounded nonlinearities. Full-block multipliers provide the complete set of QCs for such nonlinearities. However, this set is defined by an infinite number of constraints. Our paper derived a new representation for the complete set of QCs.
This construction is based on the following fact:  the set of inputs/outputs from non-repeated sector-bounded nonlinearities is equal to the set of incremental pairs from an appropriately constructed  piecewise linear function. This new characterization only requires a finite
number of matrix copositivity constraints. 
These can be implemented exactly when the
nonlinearity has dimension less than five, but copositivity relaxations are required for higher dimensions.  

\bibliographystyle{IEEEtran}
\bibliography{FBbib}

\end{document}

%% file: figures/UncSys.tex
\begin{tikzpicture}[blockdiag]
	
	\node[block,minimum width=1.3cm, minimum height=1.3cm](Plant) {$G$};
	\node[block, above= 0.4cm of Plant] (Delta) {$\Delta_\Phi$};

	\draw[<-] ($(Plant.south east)!.3!(Plant.north east)$) -- +(+1.0cm, 0) 
	node[above, name=e]{$u$};
	\draw[->] ($(Plant.south west)!.3!(Plant.north west)$) -- +(-1.0cm, 0) 
	node[above, name=d]{$y$};
	
	\draw[<-] ($(Plant.south east)!.7!(Plant.north east)$) -- +(+0.5cm, 0) |- 
	node[near start, right]{$w$} (Delta);	
	\draw[->] ($(Plant.south west)!.7!(Plant.north west)$) -- +(-0.5cm, 0) |- 
	node[near start, left ]{$v$} (Delta);
	
\end{tikzpicture}

%% file: figures/LinePlots.tex
\begin{tikzpicture}[>=stealth]
\definecolor{blue1}{RGB}{222,235,247}
\definecolor{blue2}{RGB}{158,202,225}
\definecolor{blue3}{RGB}{49,130,189}

\begin{axis}[ width = 0.5\columnwidth, height = 0.5\columnwidth, 
    name=leftplot,
    grid=none,
    xlabel=  $v$, 		
    ylabel=  $f_{\alpha\beta}(v)$, 	
    xmin = -3, xmax = 3, 
    ymin = -6, ymax = 6, 
    axis lines=middle,   
    axis line style={->},  
    xtick={-2,1.5},
    ytick={-2,4.5},
    xticklabels={$\hat{v}$,$\bar{v}$},
    yticklabels={$\hat{w}$,$\bar{w}$},         
    xticklabel shift=-17pt,
    xlabel style={at={(ticklabel* cs:1)},anchor=west},
    ylabel style={yshift = 0.6cm,anchor=north},
]

\addplot[blue2, line width = 1.25, name path=lineAlpha,domain=-10:0, samples=50,] {x};
\addplot[blue2, line width = 1.25, name path=lineBeta,domain=0:10, samples=50,] {3*x};\label{pl:pl}

\addplot[red,only marks, mark=*,mark options={scale=0.75}] coordinates {(-2,-2)};
\addplot[red,only marks, mark=*,mark options={scale=0.75}] coordinates {(1.5,4.5)};
\addplot[red, line width = 1.25, name path=lineSlope,domain=-2:1.5, samples=50,] {1.875*(x+2)-2};
\end{axis}

\begin{axis}[ width = 0.5\columnwidth, height = 0.5\columnwidth, 
 name=rightplot,  	
 at={(leftplot.east)},  
    anchor=west,           
    xshift=1.0cm,          
   xlabel=  $dv$, 		
   ylabel=  $dw$, 	
   xmin = -4, xmax = 5, 
   ymin = -8, ymax = 10, 
   axis lines=middle,        
   axis line style={->},     
   xtick=\empty,        
   ytick=\empty,        
   xtick={3.5},
   ytick={6.5},
   xticklabels={$\bar{v}-\hat{v}$},
   yticklabels={$\bar{w}-\hat{w}$},         
   xlabel style={at={(ticklabel* cs:1)},anchor=west},        
   ylabel style={yshift = 0.6cm,anchor=north},
]
    
\addplot[blue, line width = 1.25, name path=lineAlpha,domain=-10:10, samples=50,] {x};

\addplot[blue, line width = 1.25, name path=lineBeta,domain=-10:10, samples=50,] {3*x};

\addplot[
    fill=blue2,
    fill opacity=0.3,
    draw=none
] fill between[of=lineAlpha and lineBeta];\label{pl:sec}

\addplot[red,only marks, mark=*,mark options={scale=0.75}] coordinates {(3.5,6.5)};

\end{axis}
\end{tikzpicture}

%% file: figures/ResultPlot.tex
\begin{tikzpicture}[>=stealth]
\definecolor{blue1}{RGB}{222,235,247}
\definecolor{blue2}{RGB}{158,202,225}
\definecolor{blue3}{RGB}{49,130,189}

\begin{axis}[ width = 1.0\columnwidth, height = 0.6\columnwidth, 
    xlabel=  $\beta$, 		
    ylabel=  $\gamma$, 	
    xmin = 0, xmax = 1.31, 
    ymin = 0, ymax = 25, 
    ]

\addplot[blue3, dotted, line width = 1.25, name path=lineBeta] table[x expr = \thisrowno{0} ,y expr = \thisrowno{1} ,col sep=comma] {figures/Results.csv};\label{pl:Md}

\addplot[blue3, dashed, line width = 1.25, name path=lineBeta] table[x expr = \thisrowno{0} ,y expr = \thisrowno{2} ,col sep=comma] {figures/Results.csv};\label{pl:Mfc}

\addplot[blue3, line width = 1.25, name path=lineBeta] table[x expr = \thisrowno{0} ,y expr = \thisrowno{3} ,col sep=comma] {figures/Results.csv};\label{pl:Mcmp}

\draw[dashed]
  (axis cs:0.0,1.3955)
  -- node[above,pos=0.8] {$\gamma_\text{nom}$}
  (axis cs:1.4,1.3955);
\end{axis}

\end{tikzpicture}